\documentclass[12pt]{iopart}

\usepackage{graphicx}
\usepackage{iopams}
\usepackage{xspace}
\usepackage{braket}

\newcommand{\dvec}[1]{\ensuremath{\boldsymbol{#1}}\xspace}
\newcommand{\vb}{\dvec{\mathrm{b}}}
\newcommand{\ve}{\dvec{\mathrm{e}}}

\newcommand{\vl}{\dvec{\mathrm{l}}}
\newcommand{\vp}{\dvec{\mathrm{p}}}
\newcommand{\vr}{\dvec{\mathrm{r}}}
\newcommand{\vsigma}{\dvec{\sigma}}

\newcommand{\meV}{\ensuremath{\mathrm{meV}}\xspace}

\begin{document}

\author{D. S. L. Abergel$^1$}
\author{J. R. Wallbank$^2$}
\author{Xi Chen$^2$}
\author{M. Mucha-Kruczy\'nski$^3$}
\author{Vladimir I. Fal'ko$^2$}

\address{$^1$ Condensed Matter Theory Center, University of Maryland,
College Park, MD 20742, USA}

\address{$^2$ Department of Physics, Lancaster University, 
Lancaster, LA1 4YB, UK}

\address{$^3$ Department of Physics, University of Bath, Claverton Down,
Bath, BA2 7AY, UK}


\title{Infrared absorption by graphene-hBN heterostructures}

\begin{abstract}
We propose a theory of optical absorption in monolayer
graphene-hexagonal boron nitride (hBN) heterostructures.
In highly oriented heterostructures, the hBN underlay produces a
long-range moir\'e superlattice potential for the graphene electrons
which modifies the
selection rules for absorption of incoming photons in the infrared to
visible frequency range.
The details of the absorption spectrum modification depend on the
relative strength of the various symmetry-allowed couplings between the
graphene electrons and the hBN, and the resulting nature of the
reconstructed band structure. 
\end{abstract}


\maketitle

Heterostructures of graphene with hexagonal boron nitride (hBN)
feature moir\'{e} patterns \cite{dean-natnano5, xue-natmat10,
decker-nl11, yankowitz-natphys8, ponomarenko-nature497, hunt-science340}
which are the result of the slight incommensurability of the periods of
these two crystals, or of their misalignment. For Dirac electrons in
graphene, a periodic geometrical pattern in the atomic arrangement of the
two superimposed honeycomb lattices translates into a hexagonal
superlattice with the period $d\approx a/ \sqrt{\delta^2 + \theta^2}
\gg a$, determined by the lattice constants $a$ of graphene and
$(1+\delta)a$ of hBN ($\delta\approx 1.8\%$), and the misalignment angle
$\theta \ll 1$. Due to the Bragg scattering from this long-period
superlattice, Dirac electrons in graphene acquire a miniband spectrum,
which is reflected in tunneling
\cite{xue-natmat10,yankowitz-natphys8} and magnetotransport
characteristics, including the recent observation of a fractal Hofstadter
spectrum in such heterostructures subjected
to a strong magnetic field \cite{ponomarenko-nature497,
hunt-science340}.

In general, there are three characteristic types of moir\'{e} miniband
structures for graphene electrons predicted by recent theories
\cite{yankowitz-natphys8, ponomarenko-nature497, hunt-science340,
wallbank-prb87, kindermann-prb86, ortix-prb86}: miniband spectra
without a distinct separation between the lowest and other minibands;
quite exceptionally, a case where the first miniband is separated from
the next miniband by a triplet of secondary Dirac points (sDPs) in both of
the graphene valleys; 
and more generically, the case where a single sDP in the first miniband
appears in one of the two inequivalent corners of hexagonal miniature
Brollouin zone (mBZ) of moir\'{e} superlattice.
Also, a generic moir\'{e} superlattice potential violates the electron-hole
symmetry in the otherwise symmetric Dirac spectrum of graphene
electrons, making the appearance of minibands different in the valence
and conduction band sides of the graphene spectrum
\cite{wallbank-prb87}. In this paper, we investigate how these three
chracteristic moir\'{e} miniband types are reflected in the absorption
spectra in the infrared to visible optical range. It has been noticed in
recent tight-binding model studies \cite{moon-prb87, moon-arXiv1308}
that, as compared to the universal absorption coefficient $g_1 = \pi
e^2/\hbar c$ per graphene layer, the absorption of light by Dirac electrons
in twisted two-layer graphenes (which also feature a moir\'{e}
superlattice) acquires the most robust features due to the edges and van
Hove singularities of the first minibands. These are affected by both
the modulation of the density of states and the sublattice structure of the
electron Bloch states in graphene modified by the superlattice. Here, we
employ the recently proposed phenomenological theory of generic
moir\'{e} superlattices in graphene on hexagonal substrates
\cite{wallbank-prb87} to analyse the absorption spectrum of the modified
Dirac electrons, with a view to using optical transmission spectroscopy
to narrow down the parameter set describing the moir\'{e} superlattice in
graphene-hBN heterostructures.

The phenomenological model for moir\'{e} superlattices in graphene-hBN
heterostructures is described by the Hamiltonian \cite{wallbank-prb87}
\begin{eqnarray}
	\hat{H}= v \vp \cdot \vsigma  
	+ v b (u_0 f_1 + \tilde{u}_0 f_2) 
	& + \zeta  v b (u_3 f_2 + \tilde{u}_3 f_1) \sigma_3 \nonumber \\
	& + \zeta  v \left[ \vl_z \times \nabla (u_1 f_2 +\tilde{u}_1 f_1) 
	\right]  \cdot \vsigma
	\label{eq:Ham}
\end{eqnarray}
where 
\begin{equation*}
	f_1 = \sum_{m=0...5} e^{i\vb_m \cdot \vr}
	\quad\mathrm{and}\quad
	f_2 = i \sum_{m=0...5}(-1)^m e^{i\vb_m \cdot \vr}.
\end{equation*}
The reciprocal lattice vectors $\vb_{m=0,1,...,5}$ are related by
$60^\circ$ rotations, and $|\vb_m| \equiv b \approx 
\frac{4\pi}{3a}\sqrt{\delta^2 + \theta^2}$.
This Hamiltonian acts on four-component wavefunctions
$(\Psi_{AK},\Psi_{BK},\Psi_{BK'},-\Psi_{AK'})^T$ 
describing the electron amplitudes on graphene sublattices $A$ and $B$
(acted upon by Pauli matrices $\sigma_i$) and in the two principal valleys
$K$ and $K'$, accounted for by $\zeta = \pm 1$ respectively in 
\Eref{eq:Ham}.
The first term in \Eref{eq:Ham} is the unperturbed Hamiltonian of
graphene where $\vp$ is the momentum of the electron.
Among the other three contributions towards $\hat{H}$, the first
describes a simple potential modulation; the second the $A$-$B$
sublattice asymmetry, locally imposed by the substrate; 
and the third the modulation of $A$-$B$ hopping associated with a
pseudo-magnetic field.
In each of these contributions, the first and second terms inside the
round brackets respectively describe the inversion symmetric and
antisymmetric parts of the moir\'{e} perturbation.  
Here, we use the energy scale $vb$, so that $|u_i|,|\tilde u_i|\ll1$ are
dimensionless parameters. 

The inversion-symmetric perturbation in \Eref{eq:Ham} determines
a gapless miniband spectrum, with the sDP singularities either at the
edge of the first miniband, or embedded into a continuous spectrum at
higher energies, whereas the asymmetric part opens a `zero-energy' gap
$\Delta_0$ and gaps $\Delta_1$  at the secondary Dirac points in the
conduction and valence bands ($s=\pm1$ respectively), 
\begin{eqnarray}
\Delta_0 =24vb|u_{1}\tilde{u}_{0}+u_{0}\tilde{u}_{1}|, \nonumber \\
\Delta_1 = \sqrt{3} vb | \tilde u_0 + 2s\zeta \tilde u_1 
- \sqrt 3 \zeta \tilde u_3|. \nonumber
\end{eqnarray}
However, recent transport experiments \cite{ponomarenko-nature497,
hunt-science340} did not show any pronounced gap at the miniband edges,
and either no gap
\cite{ponomarenko-nature497} or a small gap \cite{hunt-science340} at
zero energy
($\Delta \sim 20\meV$), telling us that the inversion-asymmetric part of
the moir\'{e} superlattice potential is weak. This agrees with the
ansatz made in Reference \cite{wallbank-prb87} that only one out of the
two sublattices (either N or B atoms) of the honeycomb lattice of hBN
top layer dominates in the coupling with the graphene electrons, thus
making
the effective lattice of the hBN perturbation simple hexagonal and
prescribing inversion symmetry to the moir\'{e} potential.  Therefore,
in the following we assume that $|\tilde u_i| \ll |u_i|$ and neglect the
inversion asymmetric terms in the analysis of optical absorption in the
infrared-to-visible range. 

Besides the above described dominance of the inversion-symmetric part in
the moir\'{e} superlattice potential, very little is known for definite
about the values of the superlattice parameters in \Eref{eq:Ham}.
Two microscopic models, based on
either hopping between the graphene carbon atoms and the hBN atoms
\cite{kindermann-prb86}, or on
scattering of graphene electrons by the quadropole electric moments of
nitrogen atoms \cite{wallbank-prb87} predict similar a relationship
between coupling constants $u_0$, $u_1$, and $u_3$ in \Eref{eq:Ham}, 
\begin{equation}
	u_0 = \frac{\tilde v}{2 vb},\quad
	u_1 = \frac{-\tilde{v} \delta}{ vb\sqrt{\delta^2 +
	\theta^2}},\quad
	u_3 = -\frac{\sqrt{3}\tilde{v}}{2 vb}, \label{eq:model}
\end{equation}
with $0.6\meV \leq \tilde v \leq 3.4\meV$, to compare with $vb\approx 340\meV$ for $\theta =0$ and
$vb \approx 750\meV$ for $\theta = 2^\circ$. However, rather simplistic
approximations are used in these models, and one must assume much
larger values of superlattice potential parameters to relate the theory
to the recent magneto-transport data \cite{wallbank-prb87}.
This suggests that what these models show is that all three
inversion-symmetric interaction terms in the Hamiltonian in
\Eref{eq:Ham} should be taken into account in a comprehensive
phenomenological theory of moir\'{e} superlattice in graphene-hBN
heterostructures. 
Having all this in mind, the purpose of the following analysis is to
establish what characteristic features in the heterostructure absorption
spectrum can be attributed to one or another combination of moir\'{e}
parameters, with a view to narrowing down their choice based on the
combination of the transport data with the forthcoming optical studies. 
Note that the deviation of the optical absorption by
electrons in a heterostructure from the universal graphene
absorption coefficient $g_1 = \pi e^2/\hbar c$ would be most pronounced
in a spectral range around $\omega \sim vb$ (from infrared at $\theta
=0^\circ$ to visible at $\theta \sim
5^\circ$): for much lower photon frequencies, electron states are almost
the same as in the unperturbed Dirac spectrum \cite{wallbank-prb87},
whereas photons of much higher energies involve transitions between
numerous overlapping minibands such that individual spectral features would
be smeared out by the faster inelastic relaxation of photoexcited
electrons and holes.

The coefficient of absorption of light described by energy $\omega$ and polarization $\ve$ is 
\begin{equation*}
	g(\omega) = \frac{8\pi\hbar}{c\omega\mathcal{A}} \mathrm{Im}
	\sum_{\vp,\lambda,\lambda'} \frac{f_{\vp\lambda'} - f_{\vp\lambda}}%
	{\omega+\epsilon_{\vp\lambda} - \epsilon_{\vp\lambda'} + i\eta}
	M_{\alpha\beta}^{\lambda\lambda'} e_\alpha^\ast e_\beta
\end{equation*}
where $\alpha,\beta =x,y$, $\epsilon_{\vp\lambda}$ stands for the
miniband energy found by diagonalization of the Hamiltonian in
\Eref{eq:Ham}, $f_{\vp\lambda}$ are the occupation numbers,
$\mathcal{A}$ is the normalization area of the miniband plane wave
states, and $\eta$ is the broadening of the energy states (we take
$\eta= vb/200$ unless otherwise stated). 
We also find numerically the eigenstates of $H$ to calculate
the matrix elements of the current operator, 
\begin{equation*}
	M_{\alpha\beta}^{\lambda\lambda'} = 
	\bra{\vp\lambda} \hat{j}^\dagger_\alpha \ket{\vp\lambda'}
	\bra{\vp\lambda'} \hat{j}_\beta \ket{\vp\lambda}
\end{equation*}
where $j_\alpha =ev\sigma_\alpha$ are Dirac current operators. The above equation 
gives the selection rules for optical transitions between the
miniband states (we neglect the momentum transfer due to absorption of
the photon), and take into account the spin and valley degeneracy. 
The $C_3$ symmetry of the moir\'e pattern implies that there is no
dependence of $g(\omega)$ on the polarization angle of the light, and
after taking into account the fact that the two valleys in the graphene
spectrum are related by time-inversion symmetry, we conclude that the
absorption spectrum is independent of the polarization state of photons. 

\begin{figure}[tb]
	\centering
	\includegraphics[]{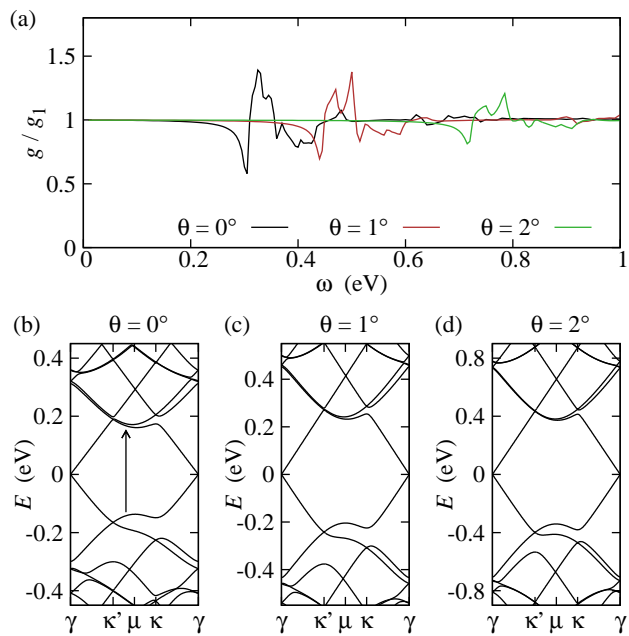}
	\caption{(a) The optical absorption spectra for the model moir\'e
	perturbation in with parameters in \Eref{eq:model} for $\tilde{v}
	= 17\meV$, $\epsilon_F = 0$, and various misalignment angles. (b)--(d) 
	Band structures corresponding to each of the spectra in (a). 
	We have marked transitions responsible for the absorption maxima in
	(b). 
	\label{fig:realpars}}
\end{figure}

\Fref{fig:realpars}(a) shows the features of the absorption spectrum
when the Fermi energy $\epsilon_F$ is at the Dirac point,
calculated for the realization of the moir\'e superlattice with
substantially sizeable amplitudes and the weight of parameters
$u_0$, $u_1$, and $u_3$ set in \Eref{eq:model}.
In this case, the electron spectrum belongs to the most generic
type: it features a sDP at the corner of the first miniband on
the valence band side and overlapping bands on the conduction band side.
It is strongly electron-hole asymmetric, which makes the spectral
features of the superlattice less pronounced. 
The optically active transitions which provide the deviation from the
standard absorption $g(\omega) = g_1$ come from the edge of the mBZ, as
shown by the arrow in \Fref{fig:realpars}(b).
This figure also shows the tendency of the
spectral features to stretch into higher energies and gradually decrease
in size with increasing misalignment angle.  

\begin{figure}[tb]
	\centering
	\includegraphics[]{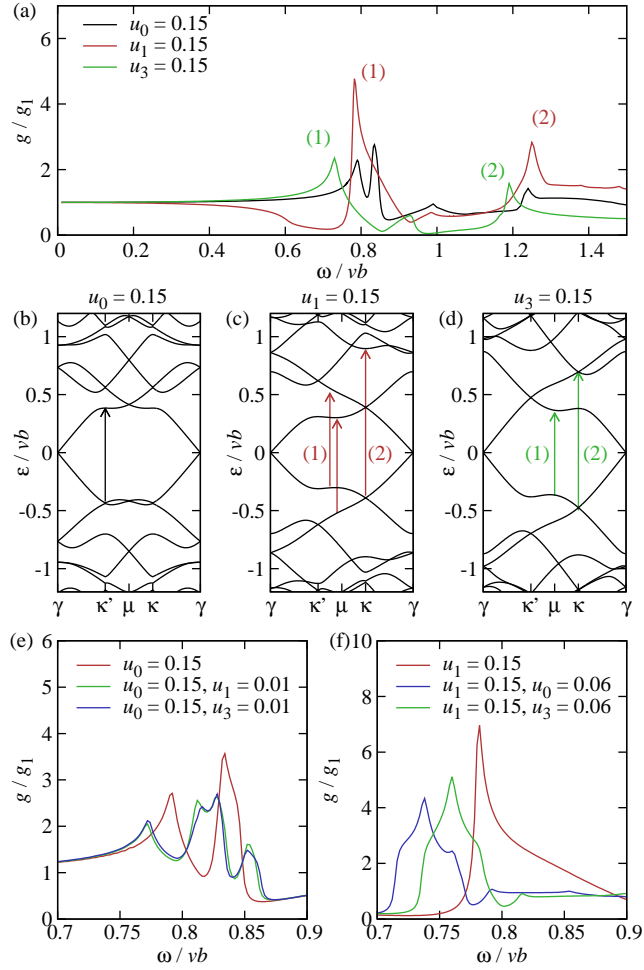}
	\caption{(a) Absorption spectra for each of the interaction terms in
	\Eref{eq:Ham} with $\epsilon_F = 0$.
	(b)--(d) The corresponding band structures with the transitions that
	make the strongest contribution to the labelled peaks in (a) marked
	with vertical arrows.
	(e) The change in the $u_0=0.15$, $u_1=u_3=0$ double peak at
	$\omega/vb \approx 0.8$ with addition of weak $u_1$ and $u_3$
	interaction terms.
	(f) The change in the $u_1=0.15$, $u_0=u_3=0$ peak at
	$\omega/vb\approx 0.8$ due to the addition of strong
	electron-hole symmetry-breaking terms $u_0$ and $u_3$.
	In (e) and (f), we have $\eta = vb/500$.
	\label{fig:spectra}}
\end{figure}

It is also instructive to analyse spectra for more peculiar realizations
of moir\'e superlattice, starting only with one of the three terms in
the perturbation, and then increasing the size of the others. The
corresponding evolution of the absorption spectra for $\epsilon_F=0$ 
is shown in \Fref{fig:spectra}(a) for each of the three interaction
terms.
The distinctive feature of a pure $u_0$ interaction [black line, band
structure shown in \Fref{fig:spectra}(b)] is the double peak
structure near $\omega/vb \approx 0.8$. \Fref{fig:spectra}(e)
shows the spectra for the same $u_0$ interaction with a small $u_1$ or
$u_3$ interaction added for the frequency interval near the double peak.
For both added interactions, each part of the double peak
is split in two destroying this simple structure. Therefore, the most
obvious identifying feature of a strong $u_0$ interaction is masked by
even weak additions of the other two interactions. The red line in
\Fref{fig:spectra}(a) is the spectrum for the $u_1=0.15$ interaction,
with the associated band structure in \Fref{fig:spectra}(c). There
are two key features to this spectrum, the first being that the
initial deviation from the standard $g(\omega) = g_1$ result for low
$\omega$ is downwards, not upwards as for the pure $u_0$ and $u_3$
interactions. The second key feature is the strong single peak at
$\omega/vb \approx 0.8$ due to the electron-hole symmetry of the $u_1$
interaction allowing van Hove singularities in both the valence and
conduction bands to contribute to the absorption simultaneously, as
indicated by the double arrow marked `(1)' in \Fref{fig:spectra}(c).
In \Fref{fig:spectra}(f) we show this peak with a strong mixture of the
$u_0$ and $u_3$ interactions [compare the
size of the perturbation to that in \Fref{fig:spectra}(e)].
For both additional interactions, the position of the peak has shifted a
little and decreased slightly in height, but the peak is
still clearly identifiable indicating that this spectral feature is
rather robust against perturbation by the other two interactions.
Finally, the $u_3 = 0.15$ interaction is shown by the green line in
\Fref{fig:spectra}(a) and the band structure in
\Fref{fig:spectra}(d). The identifying feature in this case is the
small peak followed by a large frequency range where the absorption is
suppressed substantially below the value of $g_1$.

\begin{figure}[tb]
	\centering
	\includegraphics[]{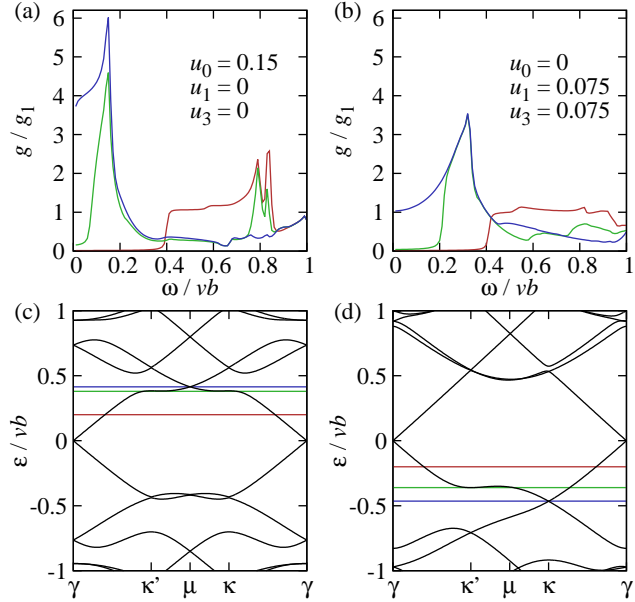}
	\caption{Variation of the optical spectrum with $\epsilon_F$. 
	(a) Optical absorption spectra when the conduction band has three
	sDPs at the mBZ edge.
	(b) Optical absorption spectra when the valence band has one sDP at
	the mBZ corner.
	(c) and (d) Band structures associated with (a) and (b),
	respectively. The horizontal coloured lines show the Fermi energy
	for each of the spectra in (a) and (b).
	\label{fig:muvary}}
\end{figure}

\Fref{fig:muvary} illustrates several examples of how the
absorption spectrum would be modified by change in the carrier density
(and Fermi energy $\epsilon_F$) in the heterostructure. In contrast to
unperturbed graphene (where Pauli blocking simply suppresses absorption
at $\omega <2\epsilon_F$) here, due to Bragg scattering of electrons by
the superlattice potential,
empty states in higher minibands of the valence band or filled
states in higher minibands of the conduction band open new absoption
channels.

In conclusion, we have demonstrated that optical spectroscopy with
infra-red and visible radiation may be used to gain insight into
the detailed characterization of the interaction between layers in
graphene-hBN heterostructures with a small misalignment angle. 
Since the exact parameters of this interaction are unknown, we have
described the general features of optical spectroscopy due to each of
the interaction terms allowed by symmetry, and linked these parameters
to the formation of secondary Dirac points in the heterostructure
spectrum and nearby van Hove singularities in the moir\'e miniband
spectra. 
We also show that the modification of the optical transitions rules, due
to the Bragg scattering of graphene electrons off the moir\'e
superlattice, modify the doping dependence of graphene absorption
spectrum, in a manner very sensitive to the detailed structure of 
moir\'e superlattice potential. 

\ack
We thank A.~Kuzmenko and K.~Novoselov for useful discussions. DSLA
acknowledges support from CMTC-LPS-NSA and US-ONR-MURI; JW was supported
by CDT NOWNANO; 
MMK acknowledges support from EPSRC First Grant EP/L013010/1;
VF thanks Royal Society Wolfson Research Merit Award,
ERC Adv Grant `Graphene and Beyond' and ERC Synergy Grant `Hetero2D'
for financial support.


\section*{References}

\begin{thebibliography}{99}

\bibitem{dean-natnano5}
	Dean CR \textit{et al.}
	2010 \textit{Nat.~Nano} \textbf{5} 722.

\bibitem{xue-natmat10}
	Xue J, Sanchez-Yamagishi J, Bulmash D, Jacquod P, Deshpande A,
	Watanabe K, Taniguchi T, Jarillo-Herrero P and Leroy BJ 2011
	\textit{Nat.~Mater.}~\textbf{10} 282.

\bibitem{yankowitz-natphys8}
	Yankowitz M, Xue J, Cormode D, Sanchez-Yamagishi JD, Watanabe K,
	Taniguchi T, Harillo-Herrero P, Jacquod P and LeRoy BJ 2012
	\textit{Nat.~Phys.} \textbf{8} 382.

\bibitem{decker-nl11}
	Decker R, Wang Y, Brar VW, Regan W, Tsai H-Z, Wu Q, Gannett W, Zettl
	A and Crommie MF 2011 \textit{Nano Letters} \textbf{11} 2291.

\bibitem{ponomarenko-nature497}
	Ponomarenko LA \textit{et al.}
	2013 \textit{Nature} \textbf{497} 594.

\bibitem{hunt-science340}
	Hunt B, \textit{et al.}
	2013 \textit{Science} \textbf{340} 1427.

\bibitem{wallbank-prb87}
	Wallbank JR, Patel AA, Mucha-Kruczy\'nski M, Geim AK and Fal'ko VI
	2013 \textit{Phys.~Rev.~}B \textbf{87} 245408.

\bibitem{kindermann-prb86}
	Kindermann M, Uchoa B and Miller DL 2012 \textit{Phys.~Rev.~}B
	\textbf{86} 115415.

\bibitem{ortix-prb86}
	Ortix C, Yang L and van den Brink J 2012 \textit{Phys.~Rev.~}B
	\textbf{86} 081405.

\bibitem{moon-prb87}
	Moon P and Koshino M 2013 \textit{Phys.~Rev.~}B \textbf{87} 205404.

\bibitem{moon-arXiv1308}
	Moon P and Koshino M 2013 Optical probing of the Hofstadter
	butterfly \textit{Preprint} arXiv:1308.0713.
\end{thebibliography}

\end{document}